# MULTIVALUED SUBSETS UNDER INFORMATION THEORY

Indraneel C Dabhade

# ABSTRACT

In the fields of finance, engineering and sciences data mining/ machine learning has held an eminent position in predictive analysis. Complex algorithms and adaptive decision models have contributed towards streamlining research as well as improve forecasting. Extensive study in areas surrounding computation and mathematical sciences has primarily been responsible for the field's development. Classification based modeling, which holds a prominent position amongst the different rule-based algorithms, is one of the most widely used decision making tool. The decision tree has a place of profound significance in classification modeling. A number of heuristics have been developed over the years to refine its decision making process. Most heuristics applied to such tree-based learning algorithms derive their roots from Shannon's 'Information Theory'. The current application of this theory is directed towards individual assessment of the attribute-values. The proposed study takes a look at the effects of combining these values with the aim to improve the 'Information Gain'. A search-based heuristic tool is applied for identifying the subsets sharing a better gain value than the ones presented in the GID3 approach. An application towards the feature selection stage of the mining process has been tested and presented with statistical analysis.

# DEDICATION & ACKNOWLEDGEMENT

I dedicate this work to Amma, Baba, my sister Swaroopa and Sir.

This work was part a thesis submitted to the Department of Industrial Engineering, Clemson University. I would like to thank Dr. Mary Beth Kurz, Dr. Anand Gramopadhye and Dr. Scott Shappell for the opportunity.

Chapter 1

# Introduction

Data mining is as a process of making meaningful conclusions from complex datasets. Fayyad (1996) refers to it as making patterns, associations, anomalies and statistically significant structures depending on the type of rule applied for class identification.

The process involves the following stages:

Data preprocessing.

Pattern recognition.

Interpreting results.

Data preprocessing provides meaning to raw data by removing noise and identifying attributes in the population. Pattern recognition identifies rules. Finally, the extracted patterns are interpreted as knowledge.

Data mining has been seen as an important analytical and predictive tool used in different sectors varying from industrial applications, marketing, and medical to achieving advances in image recognition, accident investigations and biometrics. With the increasing popularity of the World Wide Web, the field has found popularity among web developers. The application is broadly divided into 2 main categories namely Web Usage Mining (WUM) and Web Structure Mining (WSM) (Srivastava et al. 2000), (Costa and Gong 2005). WUM identifies prominent searches made by the user over the internet to recognize popular and emerging trends. With the advent of the social networking sites over the past decade, companies have been able to target specific users based on their set preferences, which resulted in an increased use of the internet as a marketing tool. Web marketers' use advanced data mining algorithms to classify the user search history and offer products and services as per the observed patterns. WHOWEDA (*W*are*HO*use of *WE*b*DA*ta) (Madria et al. 1999) is a prominent project in the field of web data mining. The project explored the use of the basic data mining architecture of links and nodes for creating a hyperlink structure of the web as an information source.

With its close proximity with the fields of machine learning and artificial intelligence, it finds extensive use in robotics. In soft computing, it makes use of tools such as fuzzy sets, neural networks and genetic algorithms on highly complex mixed mode/media datasets (Mitra et al. 2002). The dynamic natures of the decision-making process and combinatorial massive search spaces have led to the refinement and development of complex algorithms. These algorithms will be dealt in the later sections. Jang and Sun (1995) focus on evolving emotions in the decision making behavior within machines. They have tried to combine human behavior via fuzzy sets with learning structures of neural networks to create hybrid mining algorithms, namely the 'Neuro-Fuzzy' systems.

## 1.1 Background

Given below are the summaries of some of the most commonly used methods, which include Clustering, Classification, Regression and Association Rules.

*Clustering:* Clustering, also known as grouping creates sets of data that are identical in specific characteristics. This methodology is also sometimes referred to as k-means clustering (MacQueen 1967), where 'k' represents the number of clusters with each centered about a mean. The two main types of clustering techniques are partition based and hierarchical. The partition based technique creates either completely exclusive or overlapping groups of objects. It deals with creating rules based on the similarity of the attribute-value sets chosen to group cases together. The hierarchical method is associated with creating a tree of clusters based on the proximity of the near-by objects/points or association with other clusters. It is commonly referred to as ‘dendrogram’ (Forina et al. 2002), based on the way the tree is built, either divisive or agglomerative (Jain and Dubes 1988) (Kaufman and Rousseeuw 1990). Apart from these two main categories, some lesser-known techniques include grid-constraint based, scalable along with a few other algorithms dealing with custom data categorization. The literature for these practices can be found in Han and Kambler (2006).

*Classification:* Under this technique, the cases are placed in differing groups. The procedures behind this methodology create rules as per training and testing individual cases. This characteristic categorizes the practice under the supervised machine learning technique. A number of algorithms have been developed for classification based data mining. Some of them include k-Nearest Neighbor, Bayesian and Neural-Net based classifiers. For the nearest neighbor approach, the new case assumes the class of the nearest case/cluster (Cover and Hart 1967). This is different from class boundary identification by the decision tree learner, in which constrained boundaries identify the classes. KNN is one such popular memory based classification system. Bayesian classifiers use probability as a tool to identify the class for the test case. These classifiers show conditional independence among the attributes while identifying classes (Zhang 2004). They use the maximum likelihood function as a tool to identify the rule. For neural-net applications, the model, the activation function and the learning algorithm help in the pattern recognition and hence prove to be a useful utility in the process (Fausett 1994). More information on its commercial application and use can be found in Bigus (1996).

*Regression*: Regression in data mining works on the principle of predicting the class with the rule generated from the regression function. Based on the property of error reduction for pattern development, a number of rule-based algorithms have been developed, both for mining using linear and non-linear regression. CART is a well-known linear regression algorithm, whereas the Support Vector Machine (SVM) is a good example for the non-linear regression algorithm.

*Association Rules*: This technique uses decision support as a measure to weigh the relationship between attributes and establish rules to predict the classes. Some prominent work in this field has been conducted by Agarwal and Srikant (1994), one of which includes studying the purchasing patterns in supermarkets and creating 'Point of Sale' (POS) systems.

The focus of this research is on a search heuristic application for the principal algorithm to develop a decision tree. The following section introduces the features of the tree.

## 1.2 The Decision Tree

The decision tree is a classification- based prediction tool constructed after developing and learning rules within a dataset. Hence, most of the algorithms defined for decision tree generation are referred to as learning algorithms. Since the optimality for the decision making process is dependent on the accuracy in its construction, the generated rules need to be recursively trained and tested. The dataset is divided into two parts. The first part deals with the training data, used to learn the algorithm, or in technical sense, to define the rules for the remaining data. The second part is used to evaluate the quality of the rules. Apart from the mentioned approach, the k-fold cross validation, is also quite popular among researchers. For the purpose of this study, the dataset would be subjected to a test-train split. The data available in the real world vary ranging from categorical, ordinal and nominal, classified under continuous and discrete. The following section deals with handling the datasets for the tree based data mining technique.

### 1.3 Discretizing the Datasets.

A few researchers have attempted to use continuous attributes directly into the data mining algorithms. The Genetic Network Program (GNP) (Taboada et al. 2007) is one such algorithm that openly handles continuous attribute-values. Most of the other algorithms first subject the continuous data to discrete intervals prior to learning rules. A number of discretizers have been developed for this purpose. The process of discretization is subject to loss of information. Popular algorithms like Ant Miner use an external function like C4.5 (Quinlan 1992) based discretizer C4.5-Disc at the preprocessing stage for discretizing the continuous datasets. An entropy based discretization approach is then applied to the original ant miner. The ‘cAnt-Miner’ (Otero et al. 2008), which had the entropy measure as a function to discretize the continuous values, proved to be better than the C4.5–Disc algorithm in only two of the eight datasets that were used for the experiment. Fayyad and Irani (1993) introduce the multi interval discretization technique, which uses the Minimum Description Length Principle (MDLP) to achieve a supervised discretization scheme. The ‘Class Attribute Interdependence

Maximization' (CAIM) (Kurgan and Cios 2004) proved to be a better discretization tool than the entropy maximization algorithm for discretizing continuous attributes.

## 1.4 Survey of Classification Algorithms

Over the years, a number of classification algorithms have been developed, each aimed at pruning the decision making process. Some of the most prominent ones are described below. However, it is essential to first review the concept of entropy, which forms a key indicator to estimate an important measure, 'Information Gain'.

Shannon's entropy is defined as the uncertainty about the source of a message. As per Shannon (1948), it would take $\log n\,(q)$ amount of information to fully encode a sequence having a total probability $q$ when considering the most probable sequences from a derived set. This gives rise to extremities; if every message is different, the resultant is a maximum number of queries required to encode the next unknown message. Similarly, if every case from the dataset coming in for classification is identified to be different, the result is a maximum amount of information required to predict the class for the next case. For the other extremity, if the same type of message repeats for every case, no additional queries will be required to encode the next incoming message, hence no additional information would be required to predict the class for the new case. There are certain advantages that have been observed with the use of entropy as a heuristic for the decision-making process. Since it uses a log function, it provides a weight in the heuristic to make the right decision. For the purpose of measuring the information in bits, this study uses log to the base value 2.

$$\text{Entropy} = -\sum_{i=1}^{n} p_i \log_2 p_i$$

Where $p_i$ represents the probability of an event, which in this case is a message occurring from a given set of all messages (*n*). Deductions on the use of entropy as the decision heuristic take different interpretations based on the objective of the algorithm. Though the goal of the study is maximizing the information gain or conversely, minimizing entropy, there are applications that use the convex optimality of entropy maximization (Guiasu and Shenitzer 1985). Description on a few major classification algorithms is as follows.

### 1.4.1 The $A^q$ Algorithm (Michalski 1969)

This classification algorithm follows the simple (if-then) rule creation technique. The main heuristic checks for the purity i.e. the maximum number of examples covered for the class. The one problem with the use of this algorithm is that it shows less flexibility to modifications owing to its dependency on specific cases.

### 1.4.2 ID3 Algorithm (Quinlan 1983)

The Iterative Dichotomiser 3 **(**ID3) algorithm uses information gain as a measure to make decisions on training the rule followed by class prediction. Information gain is defined as the difference between the entropy needed to collect the information about a class H(T) and the entropy needed to conclude about a class given an attribute-value H(T|X).

Gain = H(T)-H(T|X)

Where,

H(T) = Entropy for probability distribution of the classes

H(T|X) = Entropy for probability distribution of the classes under the dataset partition for attribute-value X.

The algorithm recursively checks for the entropy in evaluating individual attribute-values while keeping a track of the attributes showing higher gains. The gain rankings along with the corresponding attributes decide the structure of the tree. More details can be found in Quinlan (1986). The algorithm faces problems with attributes carrying larger values, since the gain tends to favor attributes with larger set of values. Appendix C.1 provides the program used for calculating the information gain under the ID3 algorithm.

### 1.4.3 **CART Algorithm** (Breiman, Friedman, Olshen, and Stone 1984)

The Classification and Regression Tree algorithm (CART) considers 3 different splitting criteria namely the GINI, Twoing and the Ordered Twoing. All the three deal with the measure of change in impurity levels in splitting the dataset to either of the decision condition on the node. Problems associated with the CART and the ID3 led to the formation of the GID3 algorithm (Cheng et al. 1988).

### 1.4.4 **ASSISTANT Algorithm** (Kononenko, Brakto and Roskar 1984)

This algorithm follows a similar classification criterion as the ID3, but provides an improvement on the noise handling capacity. The algorithm tests each leaf node for further branching. The termination criterion is the test of reduction in classification accuracy.

### 1.4.5 **Generalized ID3 (GID3) Algorithm** (Cheng, Fayyad, Irani and Qian 1988)

The GID3 algorithm considers binary partitions of the attribute-values. The attribute is divided into two discrete subsets, one that contains the test attribute-value ($A=a_i$) and the other containing the rest of the values ($A \neq a_i$). The gains for these attributes are ranked. On the basis of the user provided threshold limit, the algorithm creates a measure to filter out the values displaying gains greater than the limit. These values are then collected together in a temporary attribute also known as a Phantom Attribute. This temporary attribute contains the values that would significantly contribute towards higher purity in classification. The procedure proves better than branching at individual values. The '*Threshold Limit*' is a user defined measure. The program for calculating the information gain under the GID3 rule is included in the Appendix C.2.

### 1.4.6 The CN2 Induction Algorithm (Clark and Niblett 1989)

The algorithm possesses properties of both the ID3 and the $A^q$ processes. It uses entropy as the criterion for creating an ordered set of rules. Such ordering is also responsible for limiting its general applicability. Using a Laplace Error Estimate as an alternative evaluation function, an unordered list of rules is derived, which along with the ordered set improve the usability of the algorithm (Clark and Boswell 1991). One common problem observed for the CN2 algorithm was with the specificity of rule selection.

### 1.4.7 GID3* Algorithm (Fayyad 1991)

A later refinement of the algorithm introduced as GID3*, provides a tear measure to automatically select the threshold level on the basis of how effectively the subset of the attribute-values manages to discretely separate the classes. A comparative performance measurement on the two algorithms shows an improvement in the following features of the decision tree:

a. Increase in the number of examples in individual final leaves.

b. Decrease in the average number of leaves.

c. Decrease in the error rate.

d. Reduction in the number of nodes.

This refined algorithm too works on the same principle of binary partitioning on individual values of the attribute. The rest of the values are grouped in a separate set.

#### 1.4.8 The Ant-Miner Algorithm (Parpinelli, Lopes and Freitas 2002)

The Ant-Miner algorithm developed on the ant colony optimization, proved to be a better suite against the CN2 algorithm considering the reduction of the number of rules and their simplification.

## 1.5 Critique of Current Research

As mentioned above, the ID3 algorithm uses entropy as a criterion to select the appropriate values for branching at the node. As part of the algorithm, branches are created at individual attribute-values followed by a comparison between the corresponding attribute gains. This algorithm suffers in problems arising from missing values/incomplete dataset. But at the same time, the use of information gain proves to be of good use to get an estimate of the attribute-value contribution in the measure of class purity (Fayyad 1991).

The key factors that affect these features of the decision tree are as follows:

a. The test conducted on the node.

b. Number of branches per node.

c. Distribution of the examples across the leaves.

d. Number of examples carried per branch of the tree.

Chapter 2

# Research Introduction

## 2.1 Area of Research

This research explores the search spaces in creating the binary partitions among the attribute-values. Consider, as an illustration a dataset for which the attributes A and B possess values (a1, a2, a3, a4, a5) and (b1, b2, b3, b4, b5) respectively. After applying the partition at the attribute node, the attribute A gets divided into two branches, picking up values (a1, a2) and (a3, a4, a5). Accordingly, the attribute B branches out to form 2 sets namely (b1, b2, b3) and (b4, b5). The increase in number of values increases the set space for creating the subsets. For the above case, the attribute A with 5 values, provides a search space of $2^5$ such partitions. To check for the right partition, the algorithm needs to execute the loop for 32 different partition combinations. Overall, if the attribute possesses '*r*' values, there are $2^r$ possible partition combinations to be evaluated for each attribute. This further has an implication on the choice of the attribute to create a node. This choice is based on the purity measure associated with the attribute. Most of the algorithms, especially the ones mentioned above, use information gain as the purity measure. This measure is a function of the attribute-value subsets. Higher gains better the classification. This is an adaptation of a problem suggested by Fayyad (1991). Though the focus is for individual attribute assessment, an important quest still remains in ranking search-based gain values for building decision trees i.e. given the order of the above attributes A and B, the challenge for the researchers is to either prove or disprove the existence of a search heuristic-based decision tree adhering to the rank order derived under the optimization criteria. The need is to develop a heuristic that would potentially perform fewer searches.

The task remains to define the binary decision vector. The measure of efficiency is the class purity.

A heuristic search approach has been adopted to identify the attribute-value subsets which provide a higher gain value than the existing GID3 algorithms. Though these values do not exceed the ID3 gains, an application towards ranking the features has been analyzed.

## 2.2 Measures

The proposed approach is subject to a measure of percentage classification errors under different classifiers. On such classifier 'ID3' has been used for this study, details for which are provided in the section 'Classifiers'.

## 2.3 Datasets

This study is aimed at using continuous datasets. The continuous data is subject to discretization. Since the proposed approach studies the effect of combining multiple attribute-values while evaluating subset performance, datasets with at least 2 unique values have been taken into consideration.

The datasets were obtained from the Machine Learning repository online (http://archive.ics.uci.edu/ml/) (Frank and Asuncion 2010). The reason for doing so was to establish a common platform to compare new models with the existing ones. The table [Table 1] briefly summarizes the features. The column 4 in Table 1 shows the range of the unique values found for individual datasets. A high value is important for this study, since it provides a higher search space for the multivalued subsets (MVS). The other key characteristics for consideration are outliers, spread and clustering.

Table 1. Dataset Characteristics

| Dataset | Instances | Attributes | Unique Values | Data Type | Missing Values |
|---|---|---|---|---|---|
| Iris | 150 | 4 | 22-43 | Fractional | No |
| Vehicle Silhouettes | 846 | 18 | 13-424 | Integer | No |

## 2.4 Classifier

### 2.4.1 Iterative Dichotomize 3 (ID3)

As explained in the previous chapter, the key measure in building the ID3 decision tree is the information gain. As per the rankings identified under feature selection, the decision tree based hierarchical rule structure provides a discrete measure to identify misclassifications on the testing set.

ID3 Classifier

1. Calculate the information gain for individual attributes.
2. Rank the attributes with increasing values of the information gain.
3. Subdivide the set of examples on the basis of the rule generated.

Key Features:

- Decision criterion is information gain.
- Prone to overfitting.

## 2.5 Testing Conditions

Since most of the programs make use of random number generators, it would be ideal to have all the random number generators follow the same stream with identical starting positions. For the purposes of this study, the Mersenne Twister is the pseudorandom number generator implemented for running the programs. Further details are provided in the section ‘Random Number Generation’. Codes were run on the Palmetto High Performance Computing (HPC) environment at Clemson University with a wall time of 50 hours per run. The discretized data was converted to an alphabetic state. It was further processed with additional set of macros.

The evaluation of the algorithms was based on the measure of classification errors. The datasets were divided into two parts in the ratio 70:30, the former representing the training data while the later the testing data.

## 2.6 Adaptive Simulated Annealing (ASA)

This section provides the information on the heuristic search tool used for the study. The goal is to identify the subset which shows higher gain. With the intention of reducing the time taken to reach the optimal value, an adaptive version of the Simulated Annealing (Talbi 2009) is being considered. The representation provided below (Algorithm A.1) has been modified for a maximization function.

Algorithm A.1: The Adaptive Simulated Annealing (ASA)

$T_o$: *Initial Temperature*
$T_{end}$: *Ending Temperature*
$Sol_{curr}$: *Current Solution*
$L_Sol_{curr}$: *Last Solution*
$E_{best}$: *Best Solution*
$\Delta$: *Change in Solution*
$L_b, L_t, F_h, F_l, I$: *Variables*

generate initial solution $Sol_{curr}$
initialize $F_l, F_h, E_{best}$ , $L_Sol_{curr}$
**begin**
    initialize $T_o, T_{end}$
    **while** $T_o > T_{end}$
    **begin**
        initialize $L_b, I, L_t$
        **while** $L_t < (L_b + I)$
        **begin**
            generate solution ( $Sol_{curr}$ )
            **if** solution< $F_l$ **then** change $F_l$
            **if** solution> = $F_h$ **then** change $F_h$
            evaluate $\Delta = Sol_{curr} - L_Sol_{curr}$

**if** Δ>0 **then** $L_Sol_{curr} = Sol_{curr}$

**if** Δ<0 **then if** $e^{\frac{\Delta}{T_o}}$ > RandomNumber **then** $L_Sol_{curr} = Sol_{curr}$

**then** $E_{best} = Sol_{curr}$

**end**

$L_t = L_b + (L_b.(1 - e)^{\frac{-(F_h - F_l)}{F_h}})$

**end**

lower $T_o$

**end**

The term $L_t = L_b + (L_b.(1 - e)^{\frac{-(F_h - F_l)}{F_h}})$ represents the adaptive equilibrium condition. With the inclusion of the gain criterion, the final version of the algorithm is interpreted below. One of the key tasks for evaluating the objective function was to define the Class Quanta Identity (CQI). The CQI represents the distribution of the attribute-values against the classes. The intention is to provide a binary divide on the unique values resulting in a partition on the dataset. There are a few other factors considered in generating a solution under the multivalued subset scheme, which are covered in the algorithm (Algorithm A.2). Appendix C.3 provides the program used for performing the 'Adaptive Simulated Annealing'. Shown below [Algorithm A.2], is the multivalued subset variant of the Adaptive Simulated Annealing.

Algorithm A.2: Multivalued subset using the Adaptive Simulated Annealing

$T_o$: *Initial Temperature*
$T_{end}$: *Ending Temperature*
$Sol_{curr}$: *Current Solution*
$L_Sol_{curr}$: *Last Solution*
$E_{best}$: *Best Solution(Highest Gain)*
Δ: *Change in Solution*
$L_b, L_t, F_h, F_l, I$: *Variables*
*CQI: Class Quanta Identity*
$E_{config}$: *Best Configuration*

generate initial solution $Sol_{curr}$
initialize $F_l, F_h, E_{best}, E_{config}$
**begin**

initialize $T_o, T_{end}$

**while** $T_o > T_{end}$

**begin**

initialize $L_b, I, L_t$

**while** $L_t < (L_b + I)$

**begin**

Binary-Rand( $n_x 1$ )

Develop CQI for the binary subsets

generate solution $Sol_{curr}$

**if** solution< $F_l$ **then** change $F_l$

**if** solution> = $F_h$ **then** change $F_h$

evaluate $\Delta = Sol_{curr} - L_Sol_{curr}$

**if** Δ>0 **then** $L_Sol_{curr} = Sol_{curr}$

**if** Δ<0 **then if** $e^{\frac{\Delta}{T_o}} > RandomNumber$ **then** $L_Sol_{curr} = Sol_{curr}$

**then** $E_{best} = Sol_{curr}$

**then** $E_{config}$= *current subsets from CQI*

**end**

$$L_t = L_b + (L_b.(1 - e)^{\frac{-(F_h - F_l)}{F_h}})$$

**end**

lower $T_o$

**end**

### 2.6.1 Random Number Generation.

One group of random binary multipliers was generated using the same stream with changing seed values. Each stream was generated of the Mersenne Twister pseudorandom number generator. The seed values provided to the random number generator played an important role to show adequate variance in the generation of the solutions. Another key observation to be noted with the implementation of this method was in utilization of the equilibrium condition. Though this condition was meant to provide self-adjustability in the speed of implementing the algorithm, it made little to no contribution towards improving a solution at the expense of time to traverse with the decreasing temperature values.

Chapter 3

## Multivalued Subset based Feature Selection

When thinking from the perspective of building a decision tree, one of the most familiar heuristic as mentioned in the earlier section is that of information gain. The previously defined algorithms (ID3, GID3 and GID3*) have used this heuristic to construct their decision trees. Though the approach of this research doesn’t define a new decision tree, it does ask the question, what if more than a single attribute-value were to be combined? How would that influence the gain? By doing so, the search space of combining the attribute-values along with reaching the objective of improving the information gain makes the problem NP-hard.

The research initially started with the intention of having to build a decision tree that would hold leaves and be tested for more than a single attribute at a time. Researchers attempting to solve the problem arrived at the conclusion of it being NP-hard. It is much more like inferring on a multicolored leaf. Though the intention is to have a tree with leaves of different color, the hardness of the problem arrives with deciding on the optimality of its structure on the basis of a multicolored leaf. A number of researchers in the field of computer science, finance, and engineering focus on the aspect of directly using classifiers to identify the rules. The increase in the number of attributes emphasizes the need to rank them as per a decision heuristic. Would one want to consider an attribute impurely dividing the classes in the decision tree or rather for that matter, any sort of classifier? What is the problem in doing so? As had been discussed earlier, the ultimate aim for any new algorithm for generating the decision tree would be to keep the number of leaves to a minimum. The reason; lesser the number of leaves, better the decision tree. Fayyad (1991) provides the proof for the same. This study applies the heuristic behind building such decision trees in the pre-classification stage. The stage commonly known as the ‘Feature Selection’ has played a vital role in a number of fields ranging from Biology (Sundaravaradan et al. 2010), Text Classification (Forman 2007), Environmental Studies (Mitrovic et al. 2009) etc. Another popular application of this tool is in the field of Reinforcement Learning (RL) (Hachiya and Sugiyama 2010). Experts believe that the next generation evolution of algorithms would need to provide added flexibility in the decision making process. When one thinks of such a process, it could either be as simple as the ‘Q learning’ technique or a more advanced form of RL with the integration of changing environmental conditions which requires a bit of dynamic programming to learn (Sutton and Barto 1998). On the whole, the basic feature selection methodology could be divided into 3 main categories (Filter, Wrapper and Embedded) (Guyon and Elisseeff 2003). Figure 1 represents the process adopted by this study to analyze the effects of the proposed heuristic feature selection method.

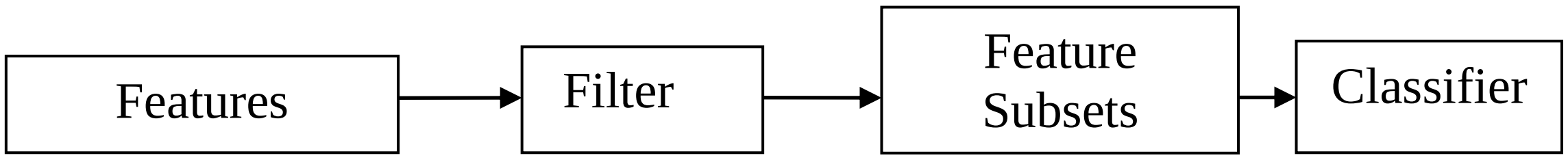

Figure 1: Filter based Feature Selection Method

The goal of the feature selection process is to identify any subset of the features satisfying the objective function. The identified subset could lie anywhere in the search path. The differentiating factor between this approach and the prior work is explained later in the section. One of the oldest feature selection criteria, information gain has found tremendous application in the commercial data mining and machine learning industry. Quite often, the filter mentioned above is the gain-based ranking method. The ranking starts with the setup of an empty scorecard. Each attribute is independently evaluated for its gain. The attributes with a higher gain occupy higher positions in the scorecard (i.e. gain-based sorting). Assume that the current classification system contains 10 attributes. An illustrative scorecard is provided in Figure 2.

| | | | | | | | | | |
|---|---|---|---|---|---|---|---|---|---|
| J | H | E | D | A | C | B | I | G | F |

| Feature Set | Classifier | Performance |
|---|---|---|
| {J, H, E, D, A, C, B, I, G, F} | ID3 | 98% |
| {J, H, E, D, A, C, B, I, G} | ID3 | 97% |
| {J, H, E, D, A, C, B, I} | ID3 | 85% |
| {J, H, E, D, A, C, B} | ID3 | 80% |
| {J, H, E, D, A, C} | ID3 | 87% |
| {J, H, E, D, A} | ID3 | 90% |
| {J, H, E, D} | ID3 | 92% |
| {J, H, E} | ID3 | 89% |
| {J, H} | ID3 | 91% |
| {J} | ID3 | 88% |

Figure 2: Illustration of a Feature Selection Process

The table above shows the performance of the varying sizes of the sets carrying the hierarchical order of the attributes being tested for a single classifier, which in this hypothetical case is the ID3 classifier. This is a case of preprocessing the attributes on a gain-based ranking, prior to testing them on the classifiers. As can be observed, the search space grows with the increase in the number of attributes. Also, the ranking heuristic works independent of the nature of the attribute-values. A number of approaches have been tested to work around the choice of attribute sets; a few contributing to the ranking based heuristics (Duch et al. 2003), while there have been approaches to use search heuristics to identify the right set, prior to classification (Vafaie and Imam 1994). The book by Liu and Motoda (2008) provides an additional insight towards the different methods used for feature selection. On the whole, the contribution of different search algorithms has been towards identifying the attribute subsets [Figure 3].

Search Space

| A | B | C | D | E | F | G | Class |
|---|---|---|---|---|---|---|---|

Figure 3: Search based Feature Selection

The ‘search space’ for this study focuses on developing the attribute-value subsets and as described in the earlier sections needs the use of a heuristic search tool. The primary goal is to find the subset providing maximum information gain. The current research does not provide a theoretical proof but is based on the assumption; attributes possessing higher gains would ultimately take a higher rank on the attribute selection scorecard. On a similar consideration, if a current attribute-value performs poorly to clearly classify a current class, would combining attribute-values increase the purity for classification? Consider the example shown in Figure 4.

| Attribute-values | Class |
|---|---|
| A.1 | 1 |
| A.1 | 1 |
| A.2 | 1 |
| A.3 | 2 |
| A.3 | 2 |
| A.3 | 3 |
| A.4 | 4 |

Figure 4: Illustration for the creation of the multivalued subsets

If attribute-value ‘A.1’ were to be tested against the rest of the attribute-value sets, it divides the class 1 in the ratio 2:1 against the rest. Hence attribute-value ‘A.1’ does not provide a pure separation for the classes. But when paired up with attribute value 'A.2', the Class '1' is distinctly separated.

### 3.1 Algorithm for the Multivalued Set

For the purposes of reducing the complexity of the representation, readers are requested to refer to the section 'Adaptive Simulated Annealing' for the framework on the search tool used in choosing the attribute-values.

Algorithm A.3: Algorithm for collecting the subsets for the maximum Information Gain.

```
begin
    size(dataset)
    for features =1:number of attributes
    begin
        calculate ClassEntropy
        perform 'Adaptive Simulated Annealing' with respect to collecting attribute –value
        pairs.
        calculate Information Gain
        collect the subsets from CQI
    end
end
```

Given below are implementations of the proposed algorithm on the datasets 'Iris' and 'Vehicle Silhouettes'. The feature subsets identified are illustrated in Figure 5 and Figure 7. The graphical comparison for the feature subset performance is shown in Figure 6 and Figure 8 respectively.

| Information Gain ID3 Ranking | | | MVS | |
|---|---|---|---|---|
| Attribute Sets | Classifier Error | | Attribute Sets | Classifier Error |
| 3,4,1,2 | 22.22% | | 4,3,1,2 | 22.22% |
| 3,4,1 | 22.22% | | 4,3,1 | 22.22% |
| 3,4 | 35.56% | | 4,3 | 35.56% |
| 3 | 35.56% | | 4 | 35.56% |

Figure 5: Figure representing the feature selection process for the dataset 'Iris'. Refer to Appendix A.2 for the nomenclature for the attribute and class names.

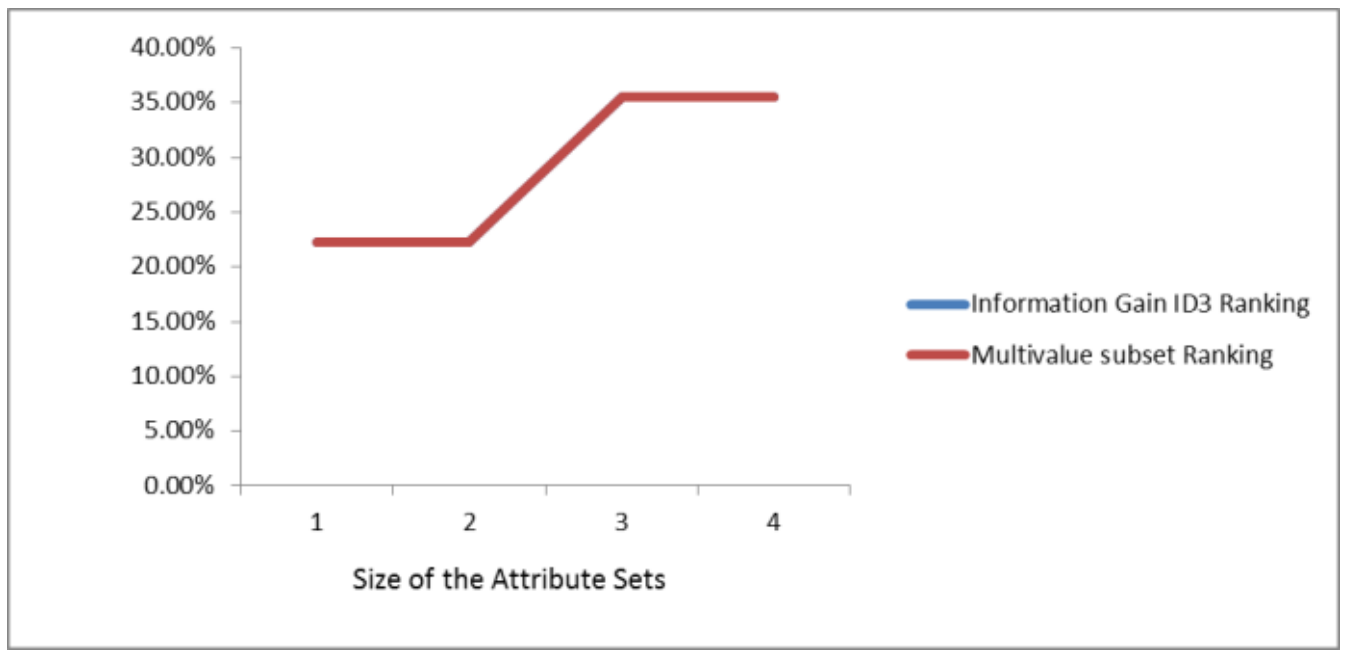

Figure 6: Classifier Error Performance between Information Gain (ID3 vs. MVS)

| Information Gain ID3 Ranking | | | | MVS Ranking | |
|---|---|---|---|---|---|
| | Classifier Error | | | | Classifier Error |
| 12,7,11,4,8,13,3,9,10,6,1,2,14,17,18,5,16,15 | 55.32% | | | 9,6,8,3,2,14,10,17,18,7,11,1,5,4,13,12,15,16 | 55.32% |
| 12,7,11,4,8,13,3,9,10,6,1,2,14,17,18,5,16 | 52.07% | | | 9,6,8,3,2,14,10,17,18,7,11,1,5,4,13,12,15 | 55.67% |
| 12,7,11,4,8,13,3,9,10,6,1,2,14,17,18,5 | 53.31% | | | 9,6,8,3,2,14,10,17,18,7,11,1,5,4,13,12 | 53.42% |
| 12,7,11,4,8,13,3,9,10,6,1,2,14,17,18 | 54.26% | | | 9,6,8,3,2,14,10,17,18,7,11,1,5,4,13 | 54.13% |
| 12,7,11,4,8,13,3,9,10,6,1,2,14,17 | 54.25% | | | 9,6,8,3,2,14,10,17,18,7,11,1,5,4 | 53.90% |
| 12,7,11,4,8,13,3,9,10,6,1,2,14 | 54.14% | | | 9,6,8,3,2,14,10,17,18,7,11,1,5 | 53.90% |
| 12,7,11,4,8,13,3,9,10,6,1,2 | 54.02% | | | 9,6,8,3,2,14,10,17,18,7,11,1 | 53.90% |
| 12,7,11,4,8,13,3,9,10,6,1 | 54.37% | | | 9,6,8,3,2,14,10,17,18,7,11 | 54.72% |
| 12,7,11,4,8,13,3,9,10,6 | 54.02% | | | 9,6,8,3,2,14,10,17,18,7 | 54.72% |
| 12,7,11,4,8,13,3,9,10 | 54.02% | | | 9,6,8,3,2,14,10,17,18 | 55.56% |
| 12,7,11,4,8,13,3,9 | 57.92% | | | 9,6,8,3,2,14,10,17 | 55.55% |
| 12,7,11,4,8,13,3 | 57.45% | | | 9,6,8,3,2,14,10 | 55.67% |
| 12,7,11,4,8,13 | 59.46% | | | 9,6,8,3,2,14 | 58.86% |
| 12,7,11,4,8 | 59.22% | | | 9,6,8,3,2 | 58.75% |
| 12,7,11,4 | 59.22% | | | 9,6,8,3 | 59.57% |
| 12,7,11 | 59.22% | | | 9,6,8 | 61.35% |
| 12,7 | 60.52% | | | 9,6 | 62.65% |
| 12 | 60.52% | | | 9 | 62.64% |

Figure 7: Figure representing the feature selection model for the dataset 'Vehicle Silhouettes'. Refer to Appendix A.2 for the nomenclature for the attribute and class names.

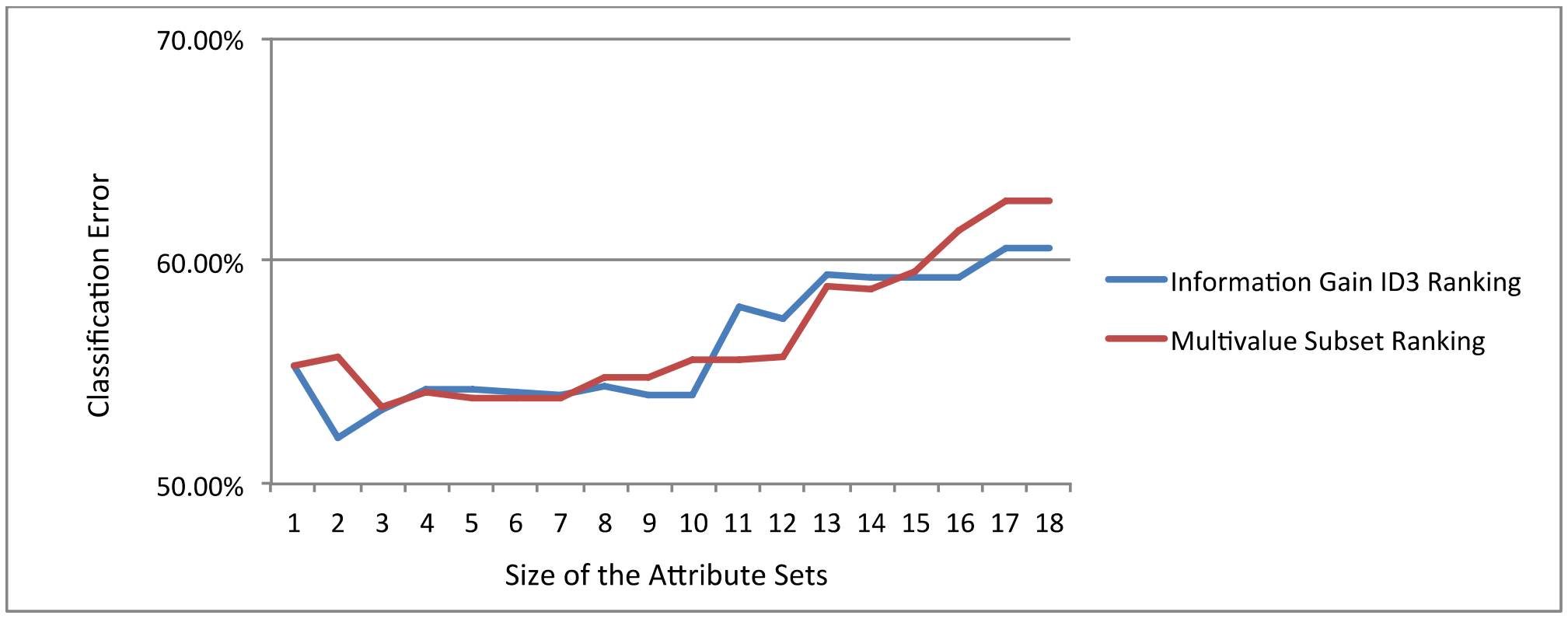

Figure 8: Vehicle Silhouettes: Classifier Error Performance between Information Gain (ID3 vs. MVS)

## 3.2 Normality Testing

The key objective of this approach is to identify the right subsets, which maximize the gain. Hence, it is essential to ensure that the heuristic displays a normal behavior while selecting the subsets. Appendix A.1 showcases the results of the normality test performed on the 'Iris' dataset. As can be observed, the normality test failed when performed on the gain values evaluated for different set sizes. But, this gain carries an upper bound. At the same time, the subset sizes selected under these gain values follow a normal distribution.

With the selected number of cluster elements (identical between the ID3 and the multivalued subsets), the algorithm did manage to find attribute sets generating lower classification errors. The results of the tests done on the dataset 'Vehicle Silhouettes' show that the attribute subsets collected as a result of the ranking provided by the multivalued subset show a lower value. Hence, if the user were to choose an attribute subset having dimensions between the maximum and the minimum value i.e. a midsize interval, the multivalued set could provide attribute sets with lower classification error in comparison to the ID3 evaluated sets.

Chapter 4

# Implications of the Work

This study has successfully managed to identify subsets of attribute value pairs which contribute towards a higher information gain. An application was made towards the feature selection process. Results indicate that the lower classification errors could be achieved for a similar sized attribute set using the MVS when compared against the traditional ID3 gain ranking method.

## 4.1 Contributions to the field of Industrial Engineering

Data mining has been viewed as a growing area of importance for its key application as a prediction tool. The above mentioned approach provides the flexibility for the data collectors to collect real time information (continuous in nature).

*Feature Selection*:

The algorithm suggested would work in identifying the right set of factors that would build a better prediction model at the cost of lowering errors on implementation samples.

# Appendices

## Appendix A.1 . Normality Tests for ‘Iris’ Dataset

### Normality Tests for Informational  Gain values:

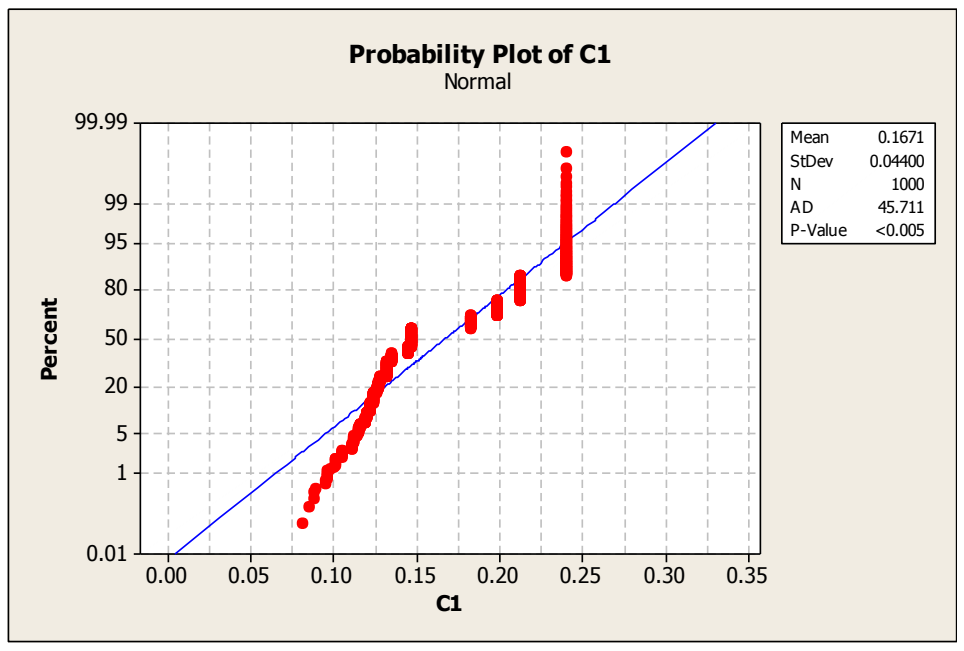

Attribute 1

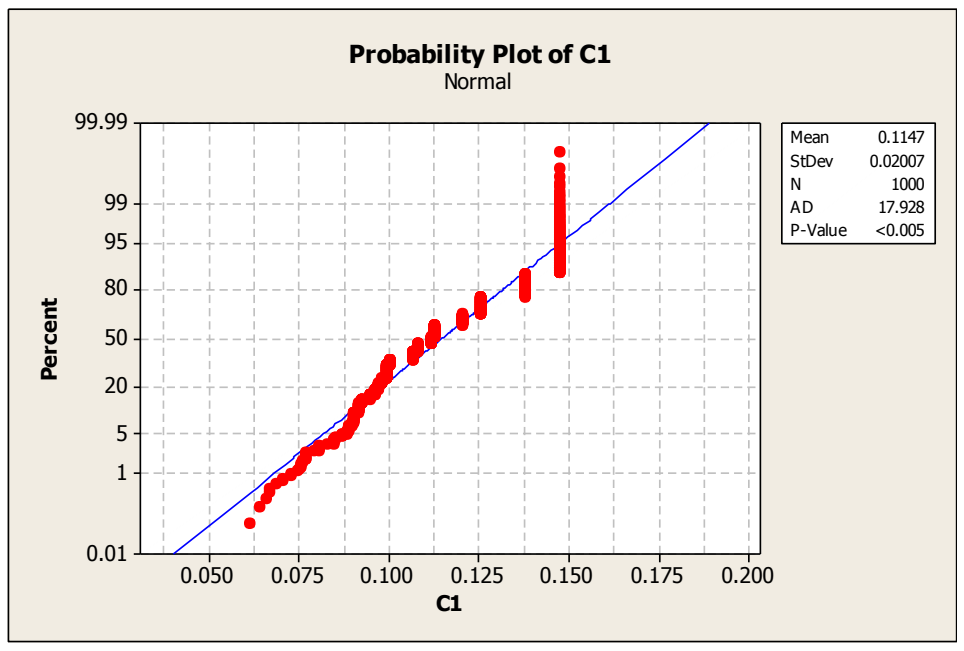

Attribute 2

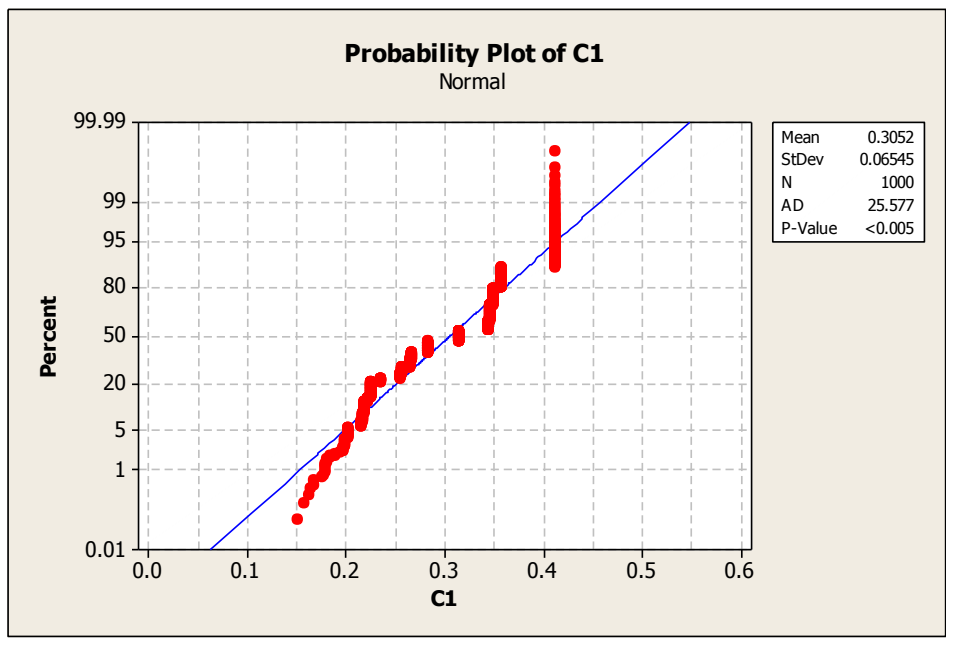

Attribute 3

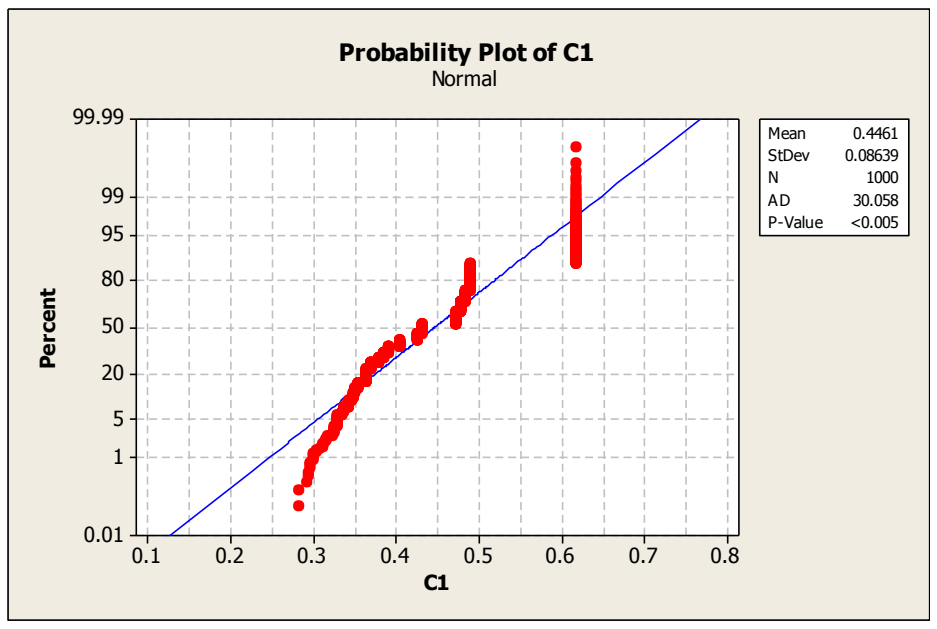

Attribute 4

### Normality Tests for Subset values:

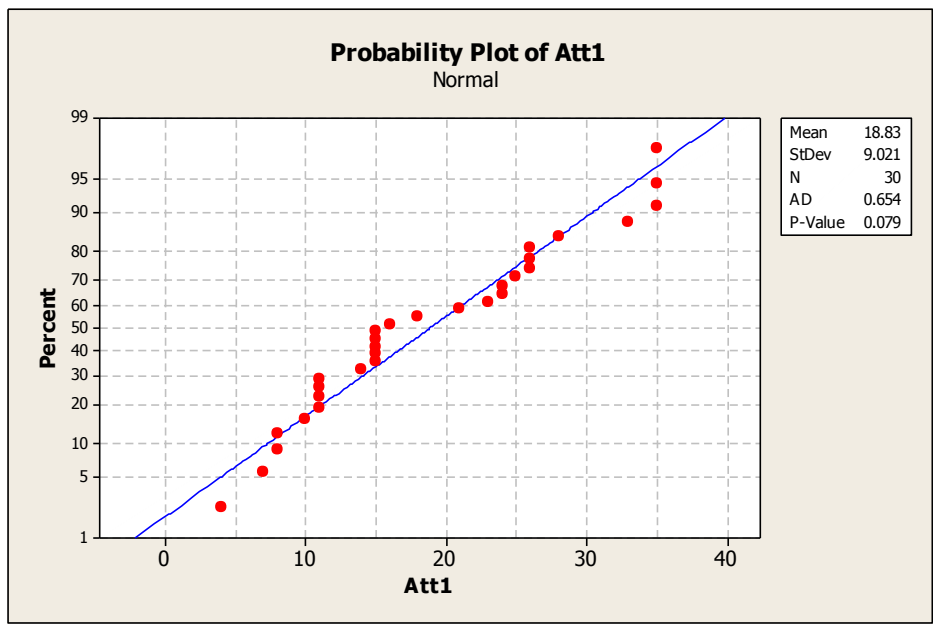

Attribute 1

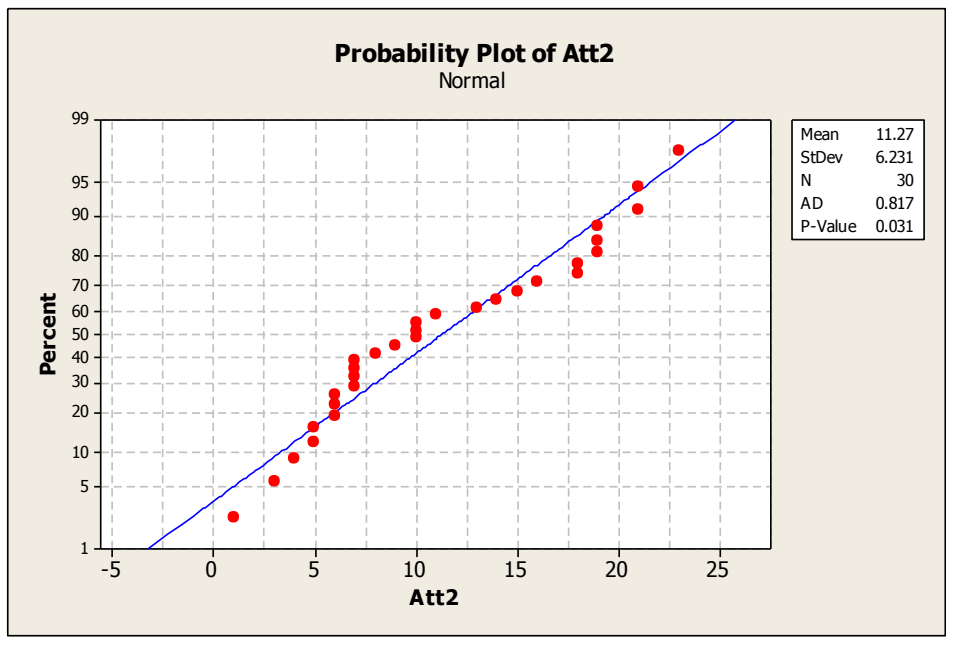

Attribute 2

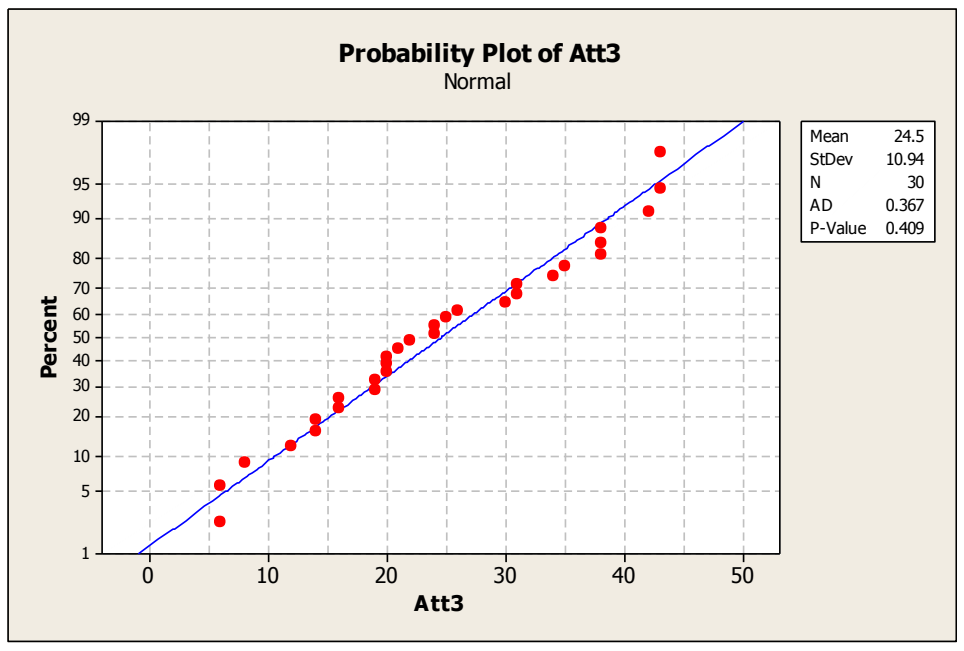

Attribute 3

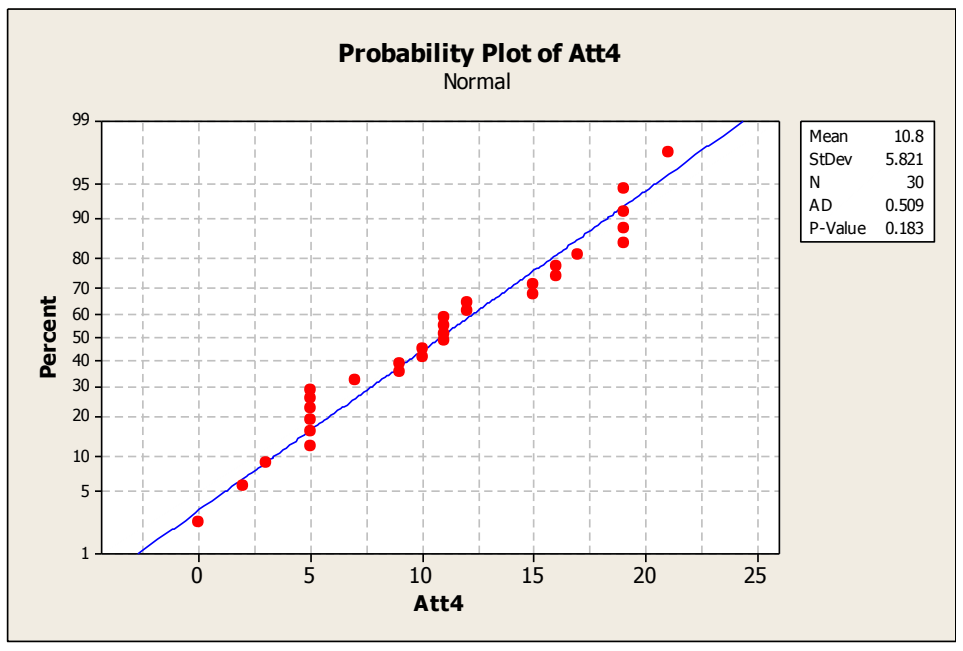

Attribute 4

## Appendix A.2 Nomenclature of the Attribute/Class Names

### Vehicle Silhouettes

| Attribute Number | Attribute Name |
|---|---|
| 1 | Compactness |
| 2 | Circularity |
| 3 | Distance Circularity |
| 4 | Radius Ratio |
| 5 | PR. Axis Aspect Ratio |
| 6 | Max. Length Aspect Ratio |
| 7 | Scatter Ratio |
| 8 | Elongatedness |
| 9 | PR. Axis Rectangularity |
| 10 | Max. Length Rectangularity |
| 11 | Scaled Variance Along Minor Axis |
| 12 | Scaled Radius Of Gyration |
| 13 | Skewness About Major Axis |
| 14 | Skewness About Minor Axis |
| 15 | Kurtosis About Major Axis |
| 16 | Hollows Ratio |
| Class 1 | OPEL |
| Class 2 | SAAB |
| Class 3 | BUS |
| Class 4 | VAN |

### Iris

| Attribute Number | Attribute Name |
|---|---|
| 1 | Sepal Length |
| 2 | Sepal Width |
| 3 | Petal Length |
| 4 | Petal Width |
| Class 1 | Iris Setosa |
| Class 2 | Iris Versicolor |
| Class 3 | Virginica |

Matlab Codes

## Appendix C.1. Program for calculating the information gain under the ID3 rule

```
close all;
clear all;
clc;
readfile = dlmread('attribute.csv', ',' , 'Range');
outfile_ID3= fopen('outfile_ID3.doc','w');
size_rows=size(readfile,1);
size_columns=size(readfile,2);
unique_class=unique(readfile(:,size_columns));
Class_entropy = zeros(numel(unique_class),1);
for class_ent=1:numel(unique_class)
    Class_entropy(class_ent) = 0-
    (length(find(readfile(:,size_columns)==(class_ent))))/size_rows.*lo
    g2((
    length(find(readfile(:,size_columns)==(class_ent))))/size_rows);
end
Entropy_class=sum(Class_entropy,1);
Sum_Final=zeros((size_columns-1),1);
for attribute = 1:(size_columns-1)
    temp_value = readfile(:,attribute);
    temp_class = readfile(:,size_columns);
    assort=[temp_value temp_class];
    assort = sortrows(assort,[1]);
    value=assort(:,1);
    class=assort(:,2);
    j = unique(value);
    ind = ones(length(j),1);
    binary_attribute= value;
    for i =1:length(j)
        ind(i) = length(find(value == j(i)));
    end
    first_value =0;
    end_value =0;
    Class_count = zeros(max(class),1);
    Class_vector = zeros(length(j), max(class));
    for i= 1:length(ind)
        first_value =end_value +1;
        end_value=first_value+(ind(i)-1);
        for window = first_value:end_value
            Class_count(class(window))= Class_count(class(window))+1;
        end
        for class_fill=1:numel(unique_class)
            Class_vector(i,class_fill)=Class_count(class_fill);
        end
        for m= 1:max(class)
            Class_count(m)=0;
        end
     end
     Sum = [j Class_vector sum(Class_vector,2)
     sum(Class_vector,2)/size_rows];
     Sum_Class=sum(Class_vector,2);
     RatioClass_vector = zeros(length(j), max(class));
     for i=1:length(j)
```

```
        for y=1:max(class)
            RatioClass_vector(i,y)=Class_vector(i,y)/Sum_Class(i);
        end
    end
    Ratio = [Sum RatioClass_vector];
    Value_Zero = RatioClass_vector;
    row_cell = ones(1,length(j));
    column_cell = ones(1, max(class));
    Cell_EliminateZero=mat2cell(Value_Zero, row_cell, column_cell);
    for p = 1: length (j)
        for t =1:max(class)
            if Cell_EliminateZero{p,t} == 0
               Cell_EliminateZero{p,t}=[];
            end
         end
     end
     log_sum=ones(length(j), 1);
     log_sum2=log_sum;
     for i =1:length(j)
         a =cell2mat(Cell_EliminateZero(i,:));
         for p = 1:length(a)
             a(p)=a(p).*log2(a(p));
         end
         log_sum(i)=0-sum(a);
         log_sum2(i)=(ind(i)/length(value)).*log_sum(i);
     end
     Sum_Final(attribute)=sum(log_sum2);
end
for att_value=1:(size_columns-1)
    fprintf(outfile_ID3,'\n%d\n',Sum_Final(att_value));
    fprintf(outfile_ID3,'\n\n\n\n--------%d-----------
    \n\n\n\n',att_value+1);
end
```

(Note: The gain is a difference in the values of the output files )

Appendix C.2. Program for calculating the information gain under the GID3 rule

```
close all;
clear all;
clc;
readfile = dlmread('attribute_Id3.csv', ',' , 'Range');
outfile= fopen('outfile.doc','w');
size_rows=size(readfile,1);
size_columns=size(readfile,2);
unique_class=unique(readfile(:,size_columns));
Class_entropy = zeros(numel(unique_class),1);
for class_ent=1:numel(unique_class)
    Class_entropy(class_ent) = 0-
    (length(find(readfile(:,size_columns)==(class_ent))))/size_rows.*log
    2((length(find(readfile(:,size_columns)==(class_ent))))/size_rows);
end
uni_max=zeros(1,size_columns-1);
for unique_max = 1:size_columns-1
    uni_max(1,unique_max)=numel(unique(readfile(:,unique_max)));
end
```

```
max_unique=max(uni_max);
Entropy_class=sum(Class_entropy,1);
temp_attribute=zeros(length(readfile),max_unique,size_columns-1);
class_attribute = zeros(length(readfile),size_columns-1);
for z = 1:size_columns-1
    readfile = sortrows(readfile,z);
    class_attribute(:,z)=readfile(:,size_columns);
    unique_value=unique(readfile(:,z));
    for y = 1:length(unique_value)
        index = find(readfile(:,z)==unique_value(y));
        for x =1:length(index)
            temp_attribute(index(x),y,z)=1;
        end
        index =0;
    end
    unique_value=0;
end
Sum_Final=zeros(max_unique,1);
for attribute = 1:size_columns-1
    for temp_uniquevalue = 1: length(unique(readfile(:,attribute)))
        temp_value = temp_attribute(:,temp_uniquevalue,attribute);
        temp_class = class_attribute(:,attribute);
        assort=[temp_value temp_class];
        assort = sortrows(assort,[1]);
        value=assort(:,1);
        class=assort(:,2);
        j = unique(value);
        ind = ones(length(j),1);
        binary_attribute= value;
        for i =1:length(j)
             ind(i) = length(find(value == j(i)));
        end
        first_value =0;
        end_value =0;
        Class_count = zeros(max(class),1);
        Class_vector = zeros(length(j), max(class));
        for i= 1:length(ind)
            first_value =end_value +1;
            end_value=first_value+(ind(i)-1);
            for window = first_value:end_value
                Class_count(class(window))=
                Class_count(class(window))+1;
            end
            for class_fill=1:numel(unique_class)
                Class_vector(i,class_fill)=Class_count(class_fill);
            end
            for m= 1:max(class)
                Class_count(m)=0;
            end
        end
        Sum = [j Class_vector sum(Class_vector,2)
        sum(Class_vector,2)/size_rows]; Sum_Class=sum(Class_vector,2);
        RatioClass_vector = zeros(length(j), max(class));
        for i=1:length(j)
            for y=1:max(class)
                RatioClass_vector(i,y)=Class_vector(i,y)/Sum_Class(i);
            end
```

```
        end
        Ratio = [Sum RatioClass_vector];
        Value_Zero = RatioClass_vector;
        row_cell = ones(1,length(j));
        column_cell = ones(1, max(class));
        Cell_EliminateZero=mat2cell(Value_Zero, row_cell, column_cell);
        for p = 1: length (j)
            for t =1:max(class)
                if Cell_EliminateZero{p,t} == 0
                   Cell_EliminateZero{p,t}=[];
                end
            end
        end
        log_sum=ones(length(j), 1);
        log_sum2=log_sum;
        for i =1:length(j)
            a =cell2mat(Cell_EliminateZero(i,:));
            for p = 1:length(a)
                a(p)=a(p).*log2(a(p));
            end
            log_sum(i)=0-sum(a);
            log_sum2(i)=(ind(i)/length(value)).*log_sum(i);
        end
        Sum_Final(temp_uniquevalue,attribute)=Entropy_class-
        sum(log_sum2);
    end
end
for att_value=1:size_columns-1
    for r=1:length(unique(readfile(:,att_value)))
        fprintf(outfile,'\n%d\n',Sum_Final(r,att_value));
    end
    fprintf(outfile,'\n\n\n\n--------%d-----------
    \n\n\n\n',att_value+1);
end
```

Appendix C.3. Program for Performing the Adaptive Simulated Annealing

```
function [Global_Best Rand_Value
NumelEl]=discretize(OriginalDataset,attribute_num)

readfile=OriginalDataset;
Dataset_Rows =size(readfile,1);
Dataset_Columns=size(readfile,2);
rand_list=rand(100000000,1);
position_rand=1;
Class_entropy = zeros(numel(unique(readfile(:,Dataset_Columns))),1);
Classes=unique(readfile(:,Dataset_Columns));
for class_count=1:numel(Classes)
    Class_entropy(Classes(class_count))=0-
    (length(find(readfile(:,Dataset_Columns)==(Classes(class_count)))))
    /Dataset_Rows.*log2(length(find(readfile(:,Dataset_Columns)==(Class
    es(class_count))))/Dataset_Rows);
end
Entropy_class=sum(Class_entropy,1);
position_rand=position_rand+1;
seed_rand=rand_list(position_rand)*1000;
```

```
[stream ]=RandStream.create('mt19937ar','NumStreams',1,'seed',seed_rand
);
random_vector=rand(stream,length(unique(readfile(:,attribute_num))),1);
constant = rand(stream,1,1);
g=random_vector>constant;
unique_value = unique(readfile(:,attribute_num));
unique_value= g.*unique(readfile(:,attribute_num));
temp_attribute=readfile(:,attribute_num);
index = find ( unique_value(:,1)>0);
j=index;
for r=1:length(index)
    j(r,1)=unique_value(index(r),1);
end
for t =1:length(j)
    index_temp=find(temp_attribute==j(t));
    for temp=1:length(index_temp)
        temp_attribute(index_temp(temp),1)=1;
    end
end
for zeros_1=1:length(readfile(:,attribute_num))
    eliminate =  find (temp_attribute(:,1)~=1);
    for k=1:length(eliminate)
        temp_attribute(eliminate(k),1)=0;
    end
end
temp_value = temp_attribute;
temp_class = readfile(:,Dataset_Columns);
assort=[temp_value temp_class];
assort = sortrows(assort,[1]);
value=assort(:,1);
class=assort(:,2);
j = unique(value);
Class_count = zeros(max(class), 1);
Class_vector = zeros(length(j), max(class));
ind = ones(length(j),1);
for i =1:length(j)
    ind(i)  = length(find(value == j(i)));
end
first_value =0;
end_value =0;
for i= 1:length(ind)
    first_value =end_value +1;
    end_value=first_value+(ind(i)-1);
    for window = first_value:end_value
        Class_count(class(window))= Class_count(class(window))+1;
    end
    for fill=1:max(Classes)
        Class_vector(i,fill)=Class_count(fill);
    end
    for m= 1:max(class)
        Class_count(m)=0;
    end
end
Sum = [j Class_vector sum(Class_vector,2)
sum(Class_vector,2)/Dataset_Rows];
Sum_Class=sum(Class_vector,2);
RatioClass_vector = zeros(length(j), max(class));
```

```
for i=1:length(j)
    for y=1:max(class)
        RatioClass_vector(i,y)=Class_vector(i,y)/Sum_Class(i);
    end
end
Ratio = [Sum RatioClass_vector];
Value_Zero = RatioClass_vector;
row_cell = ones(1,length(j));
column_cell = ones(1, max(class));
Cell_EliminateZero=mat2cell(Value_Zero, row_cell, column_cell);
for p = 1: length (j)
    for t =1:max(class)
        if Cell_EliminateZero{p,t} == 0
           Cell_EliminateZero{p,t}=[];
        end
    end
end
log_sum=ones(length(j), 1);
log_sum2=log_sum;
for i =1:length(j)
    a =cell2mat(Cell_EliminateZero(i,:));
    for p = 1:length(a)
        a(p)=a(p).*log2(a(p));
    end
    log_sum(i)=0-sum(a);
    log_sum2(i)=(ind(i)/length(value)).*log_sum(i);
end
Sum_Final=Entropy_class-sum(log_sum2);
Fh=Sum_Final;
Fl=Sum_Final;
Initial_Solution_Objective_Value=Sum_Final;
Global_Best=Initial_Solution_Objective_Value;
Rand_Value=seed_rand;
NumelEl=numel(find(g==1));
Global_Config_Best=temp_attribute;
Best_Solution=Sum_Final;
Equilibrium_Best=0;
Solution_Objective_Value=0;
T = 1000;
Tend = 1;
while T > Tend
      loop =0;
      Transition_L=2;
      Increament = 0;
      l_Transition=0;
      while l_Transition<(Transition_L+Increament)
            position_rand=position_rand+1;
            seed_rand=rand_list(position_rand)*1000;
            [stream ]= RandStream.create
            ('mt19937ar','NumStreams',1,'seed',seed_rand );
            random_vector=rand(stream,length(unique(readfile(:,attribut
            e_num))),1);
            constant = rand(stream,1,1);
            g=random_vector>constant;
            unique_value=unique(readfile(:,attribute_num));
            unique_value= g.*unique(readfile(:,attribute_num));
            temp_attribute=readfile(:,attribute_num);
```

```
index = find ( unique_value(:,1)>0);
j=index;
for r=1:length(index)
    j(r,1)=unique_value(index(r),1);
end
for t =1:length(j)
    index_temp=find(temp_attribute(:,1)==j(t));
    for temp=1:length(index_temp)
        temp_attribute(index_temp(temp),1)=1;
    end
end
for zeros_1=1:length(readfile(:,attribute_num))
     eliminate =  find (temp_attribute(:,1)~=1);
     for k=1:length(eliminate)
         temp_attribute(eliminate(k),1)=0;
     end
end
temp_value = temp_attribute(:,1);
temp_class = readfile(:,Dataset_Columns);
assort=[temp_value temp_class];
assort = sortrows(assort,[1]);
value=assort(:,1);
class=assort(:,2);
j = unique(value);
Class_count = zeros(max(class), 1);
Class_vector = zeros(length(j), max(class));
ind = ones(length(j),1);
for i =1:length(j)
    ind(i) = length(find(value == j(i)));
end
first_value =0;
end_value =0;
for i= 1:length(ind)
    first_value =end_value +1;
    end_value=first_value+(ind(i)-1);
    for window = first_value:end_value
        Class_count(class(window))=
        Class_count(class(window))+1;
    end
    for fill=1:max(Classes)
        Class_vector(i,fill)=Class_count(fill);
    end
    for m= 1:max(class)
        Class_count(m)=0;
    end
end
Sum = [j Class_vector sum(Class_vector,2)
sum(Class_vector,2)/Dataset_Rows];
Sum_Class=sum(Class_vector,2);
RatioClass_vector = zeros(length(j), max(class));
for i=1:length(j)
    for y=1:max(class)
        RatioClass_vector(i,y)=Class_vector(i,y)/Sum_Class(i
        );
    end
end
Ratio = [Sum RatioClass_vector];
```

```
Value_Zero = RatioClass_vector;
row_cell = ones(1,length(j));
column_cell = ones(1, max(class));
Cell_EliminateZero=mat2cell(Value_Zero, row_cell,
column_cell);
for p = 1: length (j)
    for t =1:max(class)
        if Cell_EliminateZero{p,t} == 0
            Cell_EliminateZero{p,t}=[];
        end
    end
end
log_sum=ones(length(j), 1);
log_sum2=log_sum;
for i =1:length(j)
    a =cell2mat(Cell_EliminateZero(i,:));
    for p = 1:length(a)
        a(p)=a(p).*log2(a(p));
    end
    log_sum(i)=0-sum(a);
    log_sum2(i)=(ind(i)/length(value)).*log_sum(i);
end
a= Entropy_class;
b=sum(log_sum2);
Sum_Final=Entropy_class-sum(log_sum2);
if Fh<Sum_Final
   Fh=Sum_Final;
end
if Fl>Sum_Final
   Fl=Sum_Final;
end
New_Solution=Sum_Final;
Best_Solution=Initial_Solution_Objective_Value;
Best_Configuration=temp_attribute;
if Initial_Solution_Objective_Value~=0;
   Solution_Objective_Value=Initial_Solution_Objective_Valu
   e;
end
Change_in_Energy = New_Solution- Solution_Objective_Value;
if Change_in_Energy>0
   Solution_Objective_Value=New_Solution;
end
if Change_in_Energy<0
   R=rand(1);
   if exp(Change_in_Energy/T)<R
      Equilibrium_Best=Best_Solution;
      Equilibrium_Config_Best=Best_Configuration;
      Move =0;
      break;
   end
   if exp(Change_in_Energy/T)>R
      Solution_Objective_Value=New_Solution;
   end
end
Best_Solution= Solution_Objective_Value;
if Equilibrium_Best<=Best_Solution
   Equilibrium_Best=Best_Solution;
```

```
                Equilibrium_Config_Best=temp_attribute;
             end
             Move =1;
             loop = loop+1;
             l_Transition=l_Transition + 1;
             Initial_Solution_Objective_Value=0;
          end
          Increament =Transition_L.*(1-exp(-(Fh-Fl)/Fh));
          Temp_Best=Equilibrium_Best;
          Temp_Config_Best=Equilibrium_Config_Best;
          if Global_Best<Temp_Best
             Global_Best=Temp_Best;
             Rand_Value=seed_rand;
             NumelEl=numel(find(g==1));
          end
          T=T.*0.90;
       end
end
```

References

Agarwal, R., and Srikant, R. (1994). Fast Algorithms for Mining Association Rules in Large Databases. Proceedings of the 20th International Conference on Very Large Data Bases, 487-499.

Bigus, P. J. (1996). Data Mining with Neural Networks:Solving Business Problems from Application Development to Decision Support. McGraw-Hill, NJ.

Breiman, L., Friedman, J.H., Olshen, R.A., and Stone, C.J. (1984). Classification and Regression Trees. Wadsworth International Group, Belmont, CA.

Cheng, J., Fayyad, U.B., Irani, K.B., and Qian, Z. (1988). Improved Decision Tree: A Generalized Version of ID3. In Proceedings of the Fifth International Conference on Machine Learning, 100-106.

Clark, P., and Boswell, R. (1991). Rule Induction with CN2: Some Recent Improvements. Machine Learning EWSL, 91, 151-163.

Clark, P., and Niblett, T. (1989). The CN2 Induction Algorithm. Machine Learning, 3, 261-283.

Costa, M.G., and Gong, Z. (2005). Web Structure Mining: An Introduction. Proceedings of the 2005 IEEE International Conference on Information Acquisition, Hong Kong and Macau, China, 590-595.

Cover, T., and Hart, P. (1967). Nearest Neighbor Pattern Classification. IEEE Transactions on Information Theory, 21-27.

Duch, W., Tomasz, W., Jacek, B., and Kachel, A. (2003). Feature Selection and Ranking Filter. Artificial Neural Networks and Neural Information Processing – ICANN/ICONIP, 251-254.

Fausett, L. (1994). Fundamentals of Neural Networks. Prentice Hall, 289-333.

Fayyad, U.M. (1991). On Induction Trees for Multiple Concept Learning, University of Michigan, Ann Arbor. (UMI Order No. GAX92-08535), Retrieved on May 20, 2010.

Fayyad, U.M., and Irani, K. (1993). Multi-Interval Discretization of Continuous-ValuedAttributes for Classification Learning. IJCAI 1993, 1022-1029.

Fayyad, U.M., Piatetsky-Shapiro, G., and Smyth, P. (1996). From Data Mining to Knowledge Discovery in Databases. AI Magazine, 17, 37-54.

Forina, M., Armanino,C., and Raggio., V. (2002). Clustering with Dendrograms on Interpretation Variables. Analytica Chemica Acta, 454, 1, 13-19.

Forman, G. (2007). Feature Selection for Text Classification. Computational Methods of Feature Selection, HP Laboratories, Palo Alto.

Frank, A. and Asuncion, A. (2010). UCI Machine Learning Repository [http://archive.ics.uci.edu/ml]. Irvine, CA: University of California, School of Information and Computer Science.

Guiasu, S., and Shenitzer, A. (1985). The Principle of Maximum Entropy. The Mathematical Intelligencer, 7, 42-48.

Guyon, I., and Elisseeff, A. (2003). An Introduction to Variable and Feature Selection. Journal of Machine Learning, 3, 1157-1182.

Hachiya, H., and Sugiyama, M. (2010). Feature Selection for Reinforcement Learning: Evaluation Implicit State-Reward Dependency via Conditional Information. ECML PKDD'10 Proceedings of the 2010 European conference on Machine Learning and Knowledge Discovery in Databases, 474-489.

Han, J., and Kambler, M. (2006). Data Mining: Concepts and Techniques. Morgan Kaufmann.

Jain, A.K., and Dubes, R.C. (1988). Algorithms for Clustering Data. Prentice Hall.

Jang, J., and Sun, C. T. (1995). Neuro-Fuzzy Modeling and Control. Proceedings of the IEEE, 83, 378-406.

Kaufman, L., and Rousseeuw, P. (1990). Finding Groups in Data: An Introduction to Cluster Analysis. Wiley 9th Edition, 342-347.

Kononenko, I., Bratko, I., and Roskar, E. (1984). Experiments in Automated Learning of Medical Diagnostic Rules. International School for the Synthesis of Expert's Knowledge Workshop.

Kurgan, L.A., and Cios, K.J. (2004). CAIM Discretization Algorithm. IEEE Transactions on Knowledge and Data Engineering, 16, 145-153.

Liu, H., and Matoda, H. (2008). Computational Methods of Feature Selection, Chapman and Hall/CRC.

MacQueen, J. (1967). Some Method for Classification and Analysis of Multivariate Observations. Fifth Berkeley Symposium on Mathematical Statistics and Probability, 281-297.

Madria, S.K., Bhowmick, S., -K, Ng. W., and Lim, E.P. (1999). Research Issues in Web Data Mining. DaWaK '99 Proceedings of the First International Conference on Data Warehousing and Knowledge Discovery, 303-312.

Michalski, R.S. (1969). On The Quasi-Minimal Solution Of The General Covering Problem. Proceedings of the 5th International Symposium of Information Processing, Yugoslavia, Bled, 125-128.

Mitra, S., Pal, S., and Mitra, P. (2002). Data mining in Soft Computing Framework: A Survey. IEEE Transactions on Neural Networks, 13, 3-14.

Mitrovic, D., Zeppelzauer, M., and Eidenberger, H. (2009). On Feature Selection in Environmental Sound Recognition. Proceedings of the 51st International Symposium ELMAR, 201-204.

Otero, F., Freitas, M., and Johnson, C. (2008). cAnt-Miner: An Ant Colony Classification Algorithm to Cope with Continuous Attributes. Lecture Notes in Computer Science, 5217, 48-59.

Parpinelli, R.S., Lopes, H., and Freitas, A. (2002). Data Mining with an Ant Colony Optimization Algorithm. IEEE Transactions on Evolutionary Computation, 6, 321-332.

Quinlan, J.R. (1983). Learning Efficient Classification Procedures and their Application to Chess End Games, Machine Learning. An Artificial Intelligence Approach, 463-482.

Quinlan, J.R. (1986). Induction of Decision Tree. Machine Learning, 1, 81-106.

Quinlan, J. R. (1992). C4.5: Programs for Machine Learning. Morgan Kaufmann Series in Machine Learning.

Shannon, C. E. (1948). A Mathematical Theory of Communication. The Bell System Technical Journal, 27, 379-423, 623-656.

Srivastava, J., Cooley, R., Deshpande, M., and Tan, P.N. (2000). Web Usage Mining: Discovery and Applicability of Usage Patterns from Web Data. ACM SIGKDD, 1, 12-23.

Sundaravaradan, N., Hossain, K., Shreedharan, V., Slotta, D., Vergara, P., Heath, L., and Ramakrishnan, N. (2010). Extracting Temporal Signatures for Comprehending Systems Biology Models. Proceedings of the 16th ACM SIGKDD International Conference on Knowledge Discovery and Data Mining, 453-462.

Sutton, S., and Barto, G. (1998). Reinforcement Learning 1: Introduction. (Adaptive Computation and Machine Learning), Chapter1.

Taboada, K., Shimada, K., Mabu, S., Hirasawa, K., and Hu, J. (2007). Association Rules Mining for Handling Continuous Attributes using Genetic Network Programming and Fuzzy Membership Functions. SICE Annual Conference, Japan, 2723 -2729.

Talbi, E., G. (2009). Metaheuristics. Wiley, New Jersey, USA.

Vafaie, H., and Imam, I. (1994). Feature Selection Methods: Genetic vs. Greedy-like Search. Proceeding of the 3rd International Fuzzy and Intelligent Control Systems Conference, Louisville, Kentuky.

Zhang, H. (2004). The Optimality of Naïve Bayes. FLAIRS Conference.